\documentstyle[12pt]{article}
\def\be{\begin{equation}}
\def\ee{\end{equation}}
\def\bc{\begin{center}}
\def\ec{\end{center}}
\def\bea{\begin{eqnarray}}
\def\eea{\end{eqnarray}}
\def\dd{\displaystyle}
\def\nn{\nonumber}
\def\cl{{\cal L}}
\def\cm{{\cal M}}
\def\cb{{\cal B}}
\def\cs{{\cal S}}
\def\cbt{\tilde{\cal B}}
\def\qaid{(Q^\alpha_i)^\dagger}
\def\qai{Q_\alpha^i}
\def\qbi{Q_\beta^i}
\def\qtai{\tilde{Q}^\alpha_i}
\def\qtaj{\tilde{Q}^\alpha_j}
\def\qtbid{(\tilde{Q}_\beta^i)^\dagger}
\def\mass{{\hat m}}
\def\la{\Lambda}

\catcode`@=11
\def\marginnote#1{}
\newcount\hour
\newcount\minute
\newtoks\amorpm
\hour=\time\divide\hour by60
\minute=\time{\multiply\hour by60 \global\advance\minute by-\hour}
\edef\standardtime{{\ifnum\hour<12 \global\amorpm={am}%
        \else\global\amorpm={pm}\advance\hour by-12 \fi
        \ifnum\hour=0 \hour=12 \fi
        \number\hour:\ifnum\minute<10 0\fi\number\minute\the\amorpm}}
\edef\militarytime{\number\hour:\ifnum\minute<10 0\fi\number\minute}
\def\draftlabel#1{{\@bsphack\if@filesw {\let\thepage\relax
   \xdef\@gtempa{\write\@auxout{\string
      \newlabel{#1}{{\@currentlabel}{\thepage}}}}}\@gtempa
   \if@nobreak \ifvmode\nobreak\fi\fi\fi\@esphack}
        \gdef\@eqnlabel{#1}}
\def\@eqnlabel{}
\def\@vacuum{}
\def\draftmarginnote#1{\marginpar{\raggedright\scriptsize\tt#1}}
\def\draft{\oddsidemargin 0.0truein
        \def\@oddfoot{\sl preliminary draft \hfil
        \rm\thepage\hfil\sl\today\quad\militarytime}
        \let\@evenfoot\@oddfoot \overfullrule 3pt
        \let\label=\draftlabel
        \let\marginnote=\draftmarginnote
   \def\@eqnnum{(\theequation)\rlap{\kern\marginparsep\tt\@eqnlabel}%
\global\let\@eqnlabel\@vacuum}  }
\catcode`@=12
\begin{document}
\begin{titlepage}
\vspace*{-1cm}
\hfill{CERN-TH/98-45}
\\
\vspace*{0.2cm}
\hfill{DFPD 98/TH/08}
\\
\vskip 1.0cm
\begin{center}
{\Large \bf A smooth massless limit for supersymmetric QCD
}
\end{center}
\vskip 0.8  cm
\begin{center}
{\large Ferruccio Feruglio}\footnote{e-mail address:
feruglio@padova.infn.it}$^,$\footnote{On leave from
Dipartimento di Fisica, Universit\`a di Padova,
I-35131 Padua, Italy}
\\
\vskip .1cm
Theory Division, CERN, CH-1211 Geneva 23, Switzerland
\\
\vskip .1cm
\end{center}
\vskip 1.0cm
\begin{abstract}
\noindent
We analyse in detail the behaviour of supersymmetric QCD with a number
of flavours $M$ smaller than the number of colours $N$, for quark
masses smaller than the dynamically generated scale $\la$.
In this regime, we find it useful to move from meson
superfields to Nambu--Goldstone-like variables.
In particular we work out the mass spectrum and
the set of decay constants that specify the interactions of the
low-energy theory. We explicitly check that masses
and decay constants have a consistent behaviour under decoupling
and that they satisfy current algebra requirements.
Finally we speculate about the massless limit. For vanishing quark
masses, and only in this case, the relation between
mesons and Nambu--Goldstone variables becomes singular.
When analysed in terms of the Nambu--Goldstone superfields,
the massless limit exhibits a spontaneous breaking
of the flavour symmetry, with massless Goldstone modes embedded
in an $M^2$-dimensional complex moduli space.
The symmetry-breaking order parameter is formally infinite,
but this has the only effect of turning off the interactions
between the chiral superfields. The massive case, for masses smaller
than $\la$, can be thought of as a perturbation around the massless
case,
with corrections that can be systematically computed in the effective
theory.
\end{abstract}
\vfill{
CERN-TH/98-45
\newline
DFPD 98/TH/08
\newline
\noindent
February 1998}
\end{titlepage}
\setcounter{footnote}{0}
\vskip2truecm
\section{Introduction}

Supersymmetric gauge theories have become an essential laboratory
to analyse non--perturbative features of relativistic quantum
field theories. This is due to several peculiar properties
of supersymmetric theories. In these theories, the
instanton calculus is much more tractable than in ordinary
quantum field theories. Moreover, the holomorphic
properties of the superpotential, combined with the asymptotic
behaviour and the symmetry  of the Wilsonian low-energy
theory, often allow a determinination of the full non-perturbative
superpotential, thus making a rigorous analysis of the
phase structure of the theory possible. In addition, the massless
spectrum of the theory is usually quite rich, and duality relations
among different theories can be conjectured by matching the
global anomalies. Finally,
extended supersymmetric theories are even more constrained, and
the full spectrum of the low-energy effective theory, including
the solitonic sector, can be determined.

Among the impressive results that have been obtained so far \cite{pes},
those concerning supersymmetric QCD (SQCD) with $M$ flavours and
$N$ colours, $M<N$, are very intriguing. The vacuum of the
theory is completely specified by the vacuum expectation
values (VEVs) of the lowest component of mesons, composite chiral
superfields that parametrize the flat directions of
the classical scalar potential in the massless case.
In the full quantum theory only the massive case seems to
provide a sensible theory. In the massless limit,
the scalar potential is unstable, with vacuum configurations
running to infinity. Although one expects that the generalized
chiral symmetry is realized in the spontaneously broken phase,
it is not clear how this phase, requiring massless chiral
multiplets associated to the Goldstone modes, is reached
as a smooth limit starting from the massive case.

Despite the difficulties associated to the strictly massless limit,
the regime of small non-vanishing masses is expected to be under
control. Owing to the large meson VEVs, the theory is weakly coupled
and we can identify both the full superpotential and the
K\"ahler potential of the low-energy effective theory.
Moreover, thanks to asymptotic freedom and to complementarity
\cite{comp},
that is the absence of a fundamental distinction between the Higgs
and the confining phases in SQCD, we can reasonably guess the
properties of the system. Light and weakly interacting quark
supermultiplets are expected to correspond to light colour-singlet
mesons characterized by weak $\sigma$-model interactions, {\it i.e.}
large decay constants. In the confining picture, spectrum and
interactions are also constrained by current algebra relations,
that were derived long ago \cite{vene}.

While the general features of the theory in this regime
are known, more specific aspects have not yet been analysed,
at least to the author's knowledge.
For very small quark masses, the perturbative corrections to
the K\"ahler potential are negligible and it becomes possible
to isolate pure non-perturbative effects in the relevant physical
quantities. It can be of interest to work out separately
the mass spectrum and the interaction terms and study
their exact dependence on the input parameters. It is also
interesting to see how the general current algebra requirements
are implemented. Moreover,
the explicit form of the low-energy effective Lagrangian may
also shed some light on the singular massless case.
The aim of the present work is to examine in detail the behaviour of
the system for quark masses much smaller than the characteristic
scale $\la$ of the theory. To keep the description of SQCD as close
as possible to the low-energy limit of ordinary QCD, we find it
convenient
to move from meson to Nambu--Goldstone superfields. For small quark
masses
the mesons fluctuate around large VEVs, while the Nambu--Goldstone
variables remain close to the origin of field space.
{}From the low-energy effective Lagrangian we can read the mass
spectrum and we can naturally characterize the interactions
in terms of a set of decay constants.
The decay constants increase when the quark masses decrease,
in agreement with the expectation that the interactions between
the relevant low-energy  degrees of freedom become weaker for
smaller quark masses.
The explicit expressions of masses and decay constants
satisfy the properties required by decoupling and by current algebra.

In the last part of this note we speculate about the massless
limit of the theory. This is a delicate point.
As long as the quark masses are non-vanishing, meson
and Nambu--Goldstone superfields are related by an analytic
transformation. The choice of variables is irrelevant, and any physical
observable remains invariant under the field redefinition.
In the massless limit, the mapping between the two sets of
variables become singular. It is well-known that the mesons
possess run-away vacua in the massless limit \cite{tvy} and the theory
has at most a cosmological interpretation \cite{wit}.
The metric of the scalar manifold is also singular for
meson VEVs running at infinity.
On the contrary, when the Nambu--Goldstone variables are used,
the metric remains regular and
the singular behaviour of the massless case can be restricted to
a single parameter, which controls the meson decay constants.
The elementary excitations of the system become massless
and non-interacting in the massless limit. The scalar potential
becomes completely flat.
The chiral symmetry is in the spontaneously broken phase, with
the expected number of chiral massless supermultiplets.
Even if formally the chiral symmetry is broken by an infinite
amount, the limit is reached smoothly in any physical computable
quantity.
For masses smaller than the SQCD scale $\la$, the massive case
can be seen as a small perturbation around the massless limit,
as in the case of ordinary QCD.

Although we have not yet found a clear-cut test
to prefer one of the two sets of variables, the present analysis
suggests that a smooth massless limit of SQCD, $M<N$,
is not inconceivable.

To discuss these points we proceed by recalling the notation and the
well-known results for SQCD, in the $M\le N$ case. Then we will
illustrate the key points in the simplest $M=1$ case. The central part
of the paper is devoted to the discussion of the general case.
Then we will come to the discussion of the massless limit
and finally we will conclude by summarizing the main results.

\section{Supersymmetric QCD}

The theory is defined by the Lagrangian:
\bea
\cl=\cl_{SYM} &+&
\int d^4\theta
\left[\qaid~ (e^V)_\alpha^\beta~ \qbi +
      \qtai~ (e^{-V})_\alpha^\beta~ \qtbid\right]  \nn\\
&-& \left[\int d^2 \theta~ \mass_j^i~ \qai \qtaj + {\rm
h.c.}\right]~~~~~,
\label{lag}
\eea
where $\qai$ and $\qtai$ $(\alpha=1,...,N; i=1,...,M)$ are chiral
superfields transforming in the $N$ and $\overline{N}$ representations
of the gauge group
$SU(N)$, respectively; $V$ is a set of vector superfields, suitably
contracted with the $SU(N)$ generators in the fundamental
representation. The first term, $\cl_{SYM}$, is the Lagrangian
for the pure super-Yang--Mills theory.
The superpotential consists just of
a mass term described by the mass matrix $\mass$.

At the classical level and in the massless case, $\mass=0$, the
vacua are distributed along flat directions, defining a classical
moduli space \cite{buc}. For $M<N$ the classical moduli space can be
parametrized
in terms of VEVs of the lowest components of $M^2$
meson superfields:
\be
\cm^i_j=\qai\qtaj~~~~~.
\label{mes0}
\ee
In the $M=N$ case the complex moduli space is $(N^2+1)$-dimensional,
the extra coordinate being provided by baryon composite superfields:
\bea
\cb=\epsilon^{{\alpha_1}...{\alpha_N}}~
Q_{\alpha_1}^1 ... Q_{\alpha_N}^N\nn\\
\cbt=\epsilon_{{\alpha_1}...{\alpha_N}}~
\tilde{Q}^{\alpha_1}_1 ... \tilde{Q}^{\alpha_N}_N
\label{bar}
\eea
subjected to the classical constraint:
\be
{\rm det}~\cm-\cb\cbt=0~~~~~.
\label{con}
\ee
At the generic point of the classical moduli space the $SU(N)$
gauge symmetry is spontaneously broken down to $SU(N-M)$ for
$M<N-1$, and completely for $M=N-1$ and $M=N$. In the massless case the
theory
exhibits a global flavour symmetry under the group
$SU(M)\times SU(M)\times U(1)_B\times U(1)_R$.
The first three factors correspond to the
usual chiral symmetry and to the baryon number, already present in the
non-supersymmetric theory. The additional
$U(1)_R$, peculiar to the supersymmetric case, is the anomaly-free
combination of the axial $U(1)$ symmetry and a global continuous
R-symmetry. The $U(1)_R$ charges of squarks, quarks and gauginos
are proportional to $M-N$, $-N$ and $M$, respectively.

When quantum corrections are included, in the massive case $\mass\ne
0$,
a number of exact results can be derived. First of all
the scalar potential of the theory vanishes at the minima and
supersymmetry is unbroken. Moreover the VEVs
$\langle\cm^i_j\rangle$ of the lowest components of the meson
superfields can be determined exactly:
\be
\langle\cm^i_j\rangle=\la^3
\left[{\rm det}\left({\mass\over \la}\right)\right]^{1\over N}
(\mass^{-1})^i_j
{}~~~~~~~~~~ (M\le N)~~~~~,
\label{vev}
\ee
where $\la$ denotes the renormalization group invariant,
dynamically generated scale of the theory. This result was first
obtained
from the Taylor--Veneziano--Yankielowicz (TVY) effective Lagrangian
\cite{tvy},
based on the superpotential:
\be
w_{TVY}=\cs\left[{\rm log}\left(\cs^{N-M} {\rm det} \cm\over
\la^{3N-M}\right)
-(N-M)\right]-{\rm tr}~(\mass\cm)~~~~~,
\label{tvy}
\ee
$\cs$ denoting a chiral superfield describing massive degrees of
freedom associated to the pure glue sector of the theory. We
recall that also the Affleck--Dine--Seiberg (ADS) superpotential
\cite{ads}
\be
w_{ADS}=(N-M)\left({\la^{3N-M}\over {\rm det}\cm}\right)^{1\over N-M}
+{\rm tr}~(\mass\cm)
\label{ads}
\ee
gives rise, for $M<N$, to the VEVs of eq. (\ref{vev}). The ADF
superpotential contains all non-perturbative corrections to the
classical
superpotential and can be related to the TVY superpotential by
integrating out
the chiral superfield $\cs$ from the TVY Lagrangian. The VEVs
$\langle\cm^i_j\rangle$
have also been reproduced via explicit instanton computations
\cite{ama}.

The massless limit can be carried out without problems for $M=N$. In
this
case the quantum moduli space along the mesonic branch has complex
dimension
$N^2-1$, as a consequence of the quantum constraint \cite{se1}:
\be
{\rm det}~\langle\cm\rangle=\la^{2N}~~~~~.
\label{qcon}
\ee
This reflects the spontaneous breaking of the global symmetry down to
the
diagonal subgroup $SU(M)\times U(1)_B\times U(1)_R$. Notice that the
subgroup $U(1)_R$ is unbroken and each complex direction in the moduli
space
contains precisely 1 Goldstone mode.

The meson VEVs of eq. (\ref{vev}) are at the origin of the difficulty
in dealing with the massless limit for $M<N$. In such a limit,
divergent
VEVs
are obtained for $\cm$, signalling an instability of the
scalar potential of the theory. By adopting the confining picture,
where the low-energy spectrum is
described by gauge invariant composite operators, a powerful constraint
is offered by a variant of the Dashen formula
\cite{vene},
relating pseudoGoldstone decay constants $F^a_i$ and masses
$m^\pi_{ij}$ to the VEV of a double commutator
\be
F^a_i m^\pi_{ij} F^b_j=\langle 0|[Q^a_5,[Q^b_5,w|_{\theta=0}]]|0\rangle
{}~~~~~.
\label{ven}
\ee
The charges $Q^a_5$ denote the $SU(M)\times SU(M)$ generators that are
spontaneously broken in
the chiral limit and $w|_{\theta=0}$ is the lowest-order
component of the superpotential. A similar formula holds also for the
$R$ charge and the related decay constant.
For equal quark masses, the right-hand side of eq. (\ref{ven})
is proportional to the product of the quark mass times the trace
of the squark condensate of eq. (\ref{vev}). If $m^\pi_{ij}$ are
linear in the quark masses, as suggested by complementarity,
then the above relation shows that the decay constants should
diverge, in the limit of vanishing quark masses,
as the square root of the meson VEVs.

All the material collected in this section is standard and well-known
and we have reported it here just to set the notation and to make
the presentation self-contained.

\section{$M=1$}

In this section we consider in detail the case $M=1$.
The global symmetry
is $U(1)_B\times U(1)_R$. There is a single meson $\cm$ and the
mass matrix $\mass$ reduces to a single mass $m$ that, without losing
generality, can be taken real and non-negative.
The global symmetry is spontaneously broken down to $U(1)_B$ by the
meson VEV.
The $U(1)_R$ symmetry is also explicitly broken by the mass term and,
for $m\ne 0$, one expects a massive chiral multiplet in the spectrum.
Here we are interested in the behaviour of the theory for a small mass
and we assume that $m$ is much smaller than the dynamically generated
scale $\la$.
{}From eq. (\ref{vev}) we can read the meson VEV
\footnote{From now on, we focus on one of the $N$ independent vacua.}:
\be
\langle\cm\rangle=\la^2~\left({\la\over m}\right)^{N-1\over N}~~~~~.
\label{vev1}
\ee
We now proceed to redefine the meson superfield according to
{}~\footnote{Variables similar to those considered here
have already been used in the literature to discuss
the massless limit of SQCD \cite{cds} and to analyse the
non-perturbative properties  of softly broken SQCD \cite{aha}.}:
\be
\cm=\la^2~\left({\la\over m}\right)^{N-1\over N}~{\rm
exp}(i\xi^0)~~~~~.
\label{red1}
\ee
in such a way that the chiral superfield $\xi^0$,
not yet conveniently normalized, has a vanishing VEV.
The lowest component of this superfield
contains the candidate Nambu--Goldstone boson for the spontaneous
breaking of the global $U(1)_R$, as can be seen by the fact that under
a $U(1)_R$ transformation it shifts by an amount proportional to the
parameter of the transformation.
To provide the correct normalization to the superfield $\xi^0$
we should consider the K\"ahler potential of the theory in the
low-energy
regime. By expressing it in terms of the superfield $\xi^0$ and by
asking for a canonical kinetic term,
we would be able to normalize $\xi^0$ appropriately.

For very large values of the meson VEVs, the original gauge theory
is in the weakly coupled regime and the K\"ahler potential can
be approximated by the classical one
\footnote{I thank L. Randall and R. Rattazzi for clarifying this point
to me. Logarithmic corrections to the K\"ahler potential
\cite{gri} become negligible for $m\ll\la$.},
projected along the direction of the meson superfields
\cite{se2,ran}:
\be
K=2\sqrt{\cm^\dagger \cm}~~~~~.
\label{kah}
\ee
In terms of the superfield
$\xi^0$, we obtain:
\be
K=2 |\la|^2~\left({|\la|\over m}\right)^{N-1\over N}
\left\vert {\rm exp}\left(i~{\xi^0\over 2}\right)\right\vert^2~~~~~.
\label{ka2}
\ee
By expanding $K$ in powers of $\xi^0$, we immediately see
that, in order to have a canonically normalized kinetic term,
the chiral superfield $\xi^0$ should be rescaled according to
\be
\xi^0\to{\sqrt{2}\xi^0\over F}
{}~~~,~~~~~~~~~~~
F=\la~\left({\la\over m}\right)^{N-1\over 2 N}
\label{dc}~~~~~.
\ee
The meson superfield now reads:
\be
\cm=F^2~{\rm exp}\left(i~{\sqrt{2}\xi^0\over F}\right)~~~~~.
\label{mes}
\ee
The full low-energy effective Lagrangian is defined by the
K\"ahler potential:
\be
K = 2 |F|^2
\left\vert {\rm exp}\left(i~{\xi^0\over \sqrt{2}
F}\right)\right\vert^2~~~~~,
\label{ka3}
\ee
and by the ADS superpotential, which takes the form:
\be
w= m F^2\left[(N-1)~ {\rm exp}\left(-i~{\sqrt{2}\over N-1} {\xi^0\over
F}\right)
+ {\rm exp}\left(i~{\sqrt{2} \xi^0\over F}\right)\right]~~~~~.
\label{weff}
\ee
The constant $F$ has clearly the meaning of the decay constant
associated to
the lowest component of $\xi^0$.
When we consider a very small mass $m\ll\la$, $F$ is
much larger than the SQCD scale $\la$. If we are interested in field
fluctuations and energy scales smaller than or comparable to $F$, we
can expand the exponentials in eqs. (\ref{ka3}) and (\ref{weff})
in powers of $\xi^0/F$, obtaining:
\be
K=|\xi^0|^2 +{i\over 2\sqrt{2}} |\xi^0|^2 \left({\xi^0\over F}-
{{\bar\xi^0}\over{\bar F}}\right)+...
\label{ka4}
\ee
\be
w= m \left({N\over N-1} (\xi^0)^2-{\sqrt{2}i\over 3}{N(N-2)\over
(N-1)^2}
{(\xi^0)^3\over F}+...\right)~~~~~.
\label{wexp}
\ee
Dots in the expression of $K$ denote analytic or anti-analytic terms
and
higher-order terms in the expansion, whereas in the expression
of $w$ they stand for an unessential constant and  higher-order
contributions.
The chiral superfield $\xi^0$ describes two neutral spin-0 particles
and a Majorana particle of spin 1/2. They have a common mass,
given by:
\be
m_{\xi^0}=
{2 N\over N-1}~ m~~~~~.
\label{mass10}
\ee
This mass is larger than twice the original quark mass.
It approaches $2 m$ in the large-$N$ limit. Notice the difference
with ordinary QCD. In that case, the squared pseudoscalar masses
scale linearly with the quark current mass and, for light flavours,
there is a substantial contribution from the quark--antiquark
condensate.
In the supersymmetric case, it is the mass of the supermultiplet
$\xi^0$
that scales linearly with the quark mass, and there is no contribution
from
the quark--antiquark condensate.

As recalled above, $m_{\xi^0}$ and $F$ should
satisfy a supersymmetric variant of the Dashen formula, which can,
for $M=1$, be cast in the following form:
\be
m_{\xi^0} F^2 = \langle 0|[Q_R,[Q_R,w|_{\theta=0}]]|0\rangle
{}~~~~~.
\label{vd1}
\ee
Both the $R$ charge $Q_R$ and the decay constant $F$ refers to a
convenient normalization of the $R$-current and its
relevant matrix element. Under a $U(1)_R$ transformation, the lowest
component of $\xi^0$ shifts by an amount proportional to $F$.
We see that the
right-hand side coincides with the double derivative of $w$, evaluated
at $\xi^0=0$, times $F^2$, in agreement with the left-hand side.

\section{$1<M<N$}

In this section we generalize the considerations of the previous
section to $M$ smaller than $N$ but otherwise
arbitrary. It is not restrictive to consider a diagonal mass matrix
$\mass$ with non-negative entries $m_1,...,m_M$.

The first step consists in performing an analytic
transformation from the meson variables $\cm^i_j$ to
Nambu--Goldstone-like variables. In a matrix notation:
\be
\cm=\la^2~\left[{\la\over ({\rm det}~\mass)^{1\over M}}
\right]^{{N-M\over N}}
\left({\rm det}~\mass\right)^{1\over M}~
\mass^{-1}~
{\rm exp}({i\xi^a T^a})
{}~~~~~.
\label{newv}
\ee
We have $M^2$ chiral superfields $\xi^a$ $(a=0,...,M^2-1)$.
The $M\times M$ matrices $T^a$ denote the generators
of $U(M)$ in the fundamental representation, normalized
for convenience according to ${\rm tr}(T^a T^b)=2\delta^{ab}$.
We identify with $T^0$ the generator equal to $\sqrt{2/M}$ times
the unit matrix.

In terms of the new variables, the VEVs of eq. (\ref{vev}) read:
\be
\langle\xi^0\rangle=\langle\xi^1\rangle=
...=\langle\xi^{M^2-1}\rangle=0~~~~~.
\label{vevs}
\ee
The ADS superpotential, expressed in terms of the new variables,
becomes:
\be
w=\la^2~\left[{\la\over ({\rm det}~\mass)^{1\over M}}
\right]^{{N-M\over N}}
\left({\rm det}~\mass\right)^{1\over M}~
\left[(N-M)~{\rm exp}\left(i~{-{\sqrt{2 M} \over N-M}\xi^0}\right)+
{\rm tr}\left({\rm exp}(i\xi^a T^a)\right)\right]
\label{weff1}
\ee
To provide canonical kinetic terms,
the Nambu--Goldstone chiral superfields should be conveniently
rescaled. This can be done by analysing the K\"ahler potential
of the theory, expressed in terms of the Nambu--Goldstone
supermultiplets. For small field and energy variations
around the large meson VEV, the full K\"ahler potential
can be approximated by the classical one:
\bea
K&=&2~{\rm tr}~\sqrt{\cm^\dagger \cm}\nn\\
 &=&2~|\la|^2~\left[{|\la|\over ({\rm det}~\mass)^{1\over M}}
\right]^{{N-M\over N}}
{\rm tr}~\left[{\rm exp}(-i{\bar\xi}^a T^a)
\left({\rm det}~\mass\right)^{2\over M}~
\mass^{-2} {\rm exp}(i\xi^a T^a)\right]^{1\over 2}
\label{ka5}
\eea
The K\"ahler potential $K$ and the superpotential $w$
would allow a full analysis of the low-energy theory.
Nevertheless, instead of dealing with the general case,
it is more instructive to work out two particular examples,
which we believe to be sufficiently representative of the behaviour
of the system.

\subsection{$\mass={\rm diag}(m,...,m)$}

The first example is characterized by a generic $M<N$, with a
degenerate mass
matrix $\mass={\rm diag}(m,...,m)$.
In this case the K\"ahler potential of eq. (\ref{ka5}) specializes
to:
\be
K=2~|\la|^2~\left({|\la|\over m}
\right)^{{N-M\over N}}~
{\rm tr}
\left[{\rm exp}(-i{\bar\xi}^a T^a)~
{\rm exp}(i\xi^a T^a)\right]^{1\over 2}~~~~~.
\label{ka6}
\ee
Because of the degeneracy of the quark mass matrix, the required
rescaling
of the Nambu--Goldstone superfields is uniform:
\be
\xi^a\to{\xi^a\over F}
{}~~,~~~~~~~~~~~~
F=\la~\left({\la\over m}\right)^{N-M\over 2 N}
\label{dc1}~~~~~.
\ee
The relation between meson and Nambu--Goldstone superfields
is:
\be
\cm=F^2~{\rm exp}\left(i~{\xi^a T^a\over F}\right)~~~~~.
\label{mes1}
\ee
The theory is described by the K\"ahler potential:
\bea
K&=&2~|F|^2~
{\rm tr}
\left[{\rm exp}\left(-i~{{\bar\xi}^a T^a\over {\bar F}}\right)
{\rm exp}\left(i~{\xi^a T^a\over F}\right)\right]^{1\over 2}\nn\\
&=&|\xi^a|^2+...~~~~~
\label{ka7}
\eea
and by the superpotential:
\bea
w&=&m F^2~
\left[(N-M)~{\rm exp}\left(-i~{\sqrt{2 M}\over N-M}{\xi^0\over
F}\right)+
{\rm tr}\left({\rm exp}\left(i~{\xi^a T^a\over
F}\right)\right)\right]\nn\\
&=&-m~\left[\left({N\over N-M}\right) (\xi^0)^2+(\xi^1)^2+...+
(\xi^{M^2-1})^2+...\right]~~~~~.
\label{weff2}
\eea
In the above expressions we have also listed the first relevant terms
in the expansion in powers of $\xi^a/ F$. All the interaction terms
are down by powers of $1/F$ compared to the quadratic part.
The spectrum consists of one chiral supermultiplet with mass
\be
m_{\xi^0}=
{2 N\over N-M}~ m
\label{mass0}
\ee
and $M^2-1$ degenerate supermultiplets with masses
\be
m_{\xi^1}=...=m_{\xi^{M^2-1}}=2 m~~~~~.
\label{mass}
\ee
Concerning
the scaling of the masses with $m$, the same remarks as made in the
previous section apply.
Notice that $M^2-1$ superfields have a common mass that is exactly
twice
the quark mass. The mass of the remaining superfield approaches
this value in the large-$N$ limit. As a curiosity, we note that,
although this analysis does not cover the case $M=N$, if we naively
try to extrapolate the spectrum to that case, we see that the mass
of the singlet diverges, in qualitative agreement with the fact that,
for $M=N$, only $N^2-1$ chiral superfields remain massless along
the mesonic branch.

It can easily be checked that also in this case the Dashen--Veneziano
relation, eq. (\ref{ven}), is satisfied. Finally, notice that
in the large-$N$ limit the anomaly-free $U(1)_R$ symmetry
tends to the axial $U(1)_A$ and the global flavour symmetry is promoted
to $U(M)\times U(M)$. In the case under discussion, this symmetry
is broken by the mass matrix down to the diagonal $U(M)$. We expect
$M^2$ degenerate pseudoGoldstone multiplets, as indeed can be seen
by sending $N$ to infinity in eq. (\ref{mass0}).

\subsection{$M=2$}

As a further example we consider in detail the case $M=2$,
for a general mass matrix $\mass={\rm diag}(m_1,m_2)$.
After some algebra, the quadratic part of the K\"ahler potential
given in eq. (\ref{ka5}), and specialized to $M=2$, reads:
\bea
K&=& |\la|^2~\left({|\la|\over \sqrt{m_1 m_2}}
\right)^{{N-2\over N}}~
\left[
\sqrt{m_1\over m_2}~\left\vert{\xi^0+
\xi^3\over\sqrt{2}}\right\vert^2+
\sqrt{m_2\over m_1}~\left\vert{\xi^0-
\xi^3\over\sqrt{2}}\right\vert^2\right.\nn\\
&+&
\left.{2 m_1\over m_1+m_2}~\sqrt{m_1\over m_2}~
\left\vert{\xi^1+i\xi^2\over\sqrt{2}}\right\vert^2+
{2 m_2\over m_1+m_2}~\sqrt{m_2\over m_1}~
\left\vert{\xi^1-i\xi^2\over\sqrt{2}}\right\vert^2
+...\right]~~~~~.
\label{k8}
\eea
As before, dots denote purely analytic or anti-analytic terms.
To recover canonical kinetic terms, a common rescaling of the fields is
no
longer sufficient, and several decay constants will appear, as
expected,
since the masses are not degenerate. We first rotate the
Nambu--Goldstone
fields to the natural combinations:
\be
\eta^{\pm}={\xi^0\pm\xi^3\over\sqrt{2}}~~~,~~~~~~~~~~
\xi^{\pm} ={\xi^1\pm i\xi^2\over\sqrt{2}}~~~~~.
\label{new}
\ee
Then we perform the rescaling:
\bea
\eta^+\to {\eta^+\over F_{\eta^+}},~~~~~~~~~~~~~~~
F_{\eta^+}&=&\la~
\left[{\la\over
m_1^{-{1\over N-2}} m_2^{N-1\over N-2}}
\right]^{N-2\over 2N}\nn\\
\eta^-\to {\eta^-\over F_{\eta^-}},~~~~~~~~~~~~~~~
F_{\eta^-}&=&\la~
\left[{\la\over
m_1^{N-1\over N-2} m_2^{-{1\over N-2}}}
\right]^{N-2\over 2N}\nn\\
\xi^+\to {\xi^+\over F_{\xi^+}},~~~~~~~~~~~~~~~
F_{\xi^+}&=&\la~
\left[{ \la\over
m_1^{-{N+1\over N-2}} m_2^{N-1\over N-2} ({\dd{m_1+m_2\over
2}})^{N\over N-2}}
\right]^{N-2\over 2N}\nn\\
\xi^-\to {\xi^-\over F_{\xi^-}},~~~~~~~~~~~~~~~
F_{\xi^-}&=&\la~
\left[{ \la\over
m_1^{N-1\over N-2} m_2^{-{N+1\over N-2}} ({\dd{m_1+m_2\over
2}})^{N\over N-2}}
\right]^{N-2\over 2N}~~~~~.
\label{decs}
\eea
The kinetic terms of the rescaled variables are canonical.
The quadratic part of the superpotential is given by:
\bea
w&=&-\left[
(m_1+m_2)~\xi^+\xi^- +\right.\nn\\
& &
\left.\left({N-1\over N-2}\right) m_2~ (\eta^+)^2+
\left({N-1\over N-2}\right) m_1~ (\eta^-)^2+
\left({2\over N-2}\right) \sqrt{m_1 m_2}~ \eta^+\eta^-
\right]~~~~~.
\label{weff3}
\eea
The masses of the physical states can be readily obtained from this
expression.
As a first check, when $m_1=m_2=m$, we can verify that the spectrum
consists of three degenerate supermultiplets of mass $2 m$ plus
a supermultiplet of mass $2 m N/(N-2)$, as required by eqs.
(\ref{mass0}) and (\ref{mass}). Also the four decay constants
become equal and reproduce the value of eq. (\ref{dc1}).

An important, additional check of the present picture is provided
by the limit $m_1\ll m_2$~
\footnote{Both masses are kept not larger than $\la$.}.
In this case, at energies much smaller than $m_2$, the theory should
look like
SQCD with a single flavour. Indeed, in this limit, three of the four
chiral supermultiplets have masses growing with $m_2$, whereas the last
supermultiplet, essentially given by $\eta^-$, has a mass
$ 2 m_1 N/N-1$, as can be checked from eq. (\ref{weff3}), by
including the first-order correction to the lowest-order eigenvalue.
Moreover, the decay constant $F_{\eta^-}$ of the light supermultiplet
exactly reproduces that of the $M=1$ case. To appreciate this
fact, we should remember that the renormalization group-invariant
scales
of the theories with one and two flavours are related by:
\be
\la_{(1)}^{3N-1}=\la^{3N-2}~m_2~~~~~,
\label{mat}
\ee
when the matching between the two theories is made at the scale $m_2$.
Taking into account this relation, the decay constant $F_{\eta^-}$
can be expressed as
\be
F_{\eta^-}=\la_{(1)}\left({\la_{(1)}\over m_1}\right)^{N-1\over
N}~~~~~,
\label{dc21}
\ee
exactly as required by the $M=1$ case.
We stress that this fundamental
property does not hold for an arbitrary choice of the K\"ahler
potential $K$. This was probably at the origin of the early
difficulties in discussing the regime of small quark masses.

\section{The massless limit}

At this point it is appropriate to make some comment on the massless
case.
In ordinary QCD, the massless limit
is smooth. The flavour symmetry is spontaneously broken
and massless Nambu--Goldstone bosons appear. The decay constant
of the pseudoscalar mesons is of the order of the QCD scale.
When small non-vanishing quark masses are turned on,
the theory can be regarded as a small perturbation
around the massless case and corrections can be systematically
included order by order in the symmetry-breaking parameters.
In SQCD, $M<N$, the lowest components of the meson superfields
have a scalar potential with run-away vacua in the massless limit.
This seems to prevent from the very beginning the construction
of the quantum theory.
The fate of the global symmetries in the massless limit then
becomes an ill-defined question. Apparently, the massive case
can no longer be regarded as a perturbation around the massless limit.

We may expect to detect some signals associated to this
singular limit already when the quark masses are non-vanishing
but very small. What emerges, instead, from the present analysis,
is that the
massless limit, whether it exists or not as a quantum theory,
is approached smoothly.
We recall that throughout this analysis
we have made use of the Nambu--Goldstone variables, introduced
with the purpose of being as close as possible to the standard
description of low-energy QCD.
For any non-vanishing value of $m$ the Nambu--Goldstone
variables are related to the meson variables by an analytic
transformation, explicitly given in eqs. (\ref{mes}), (\ref{newv}) and
(\ref{mes1}).
While the Nambu--Goldstone superfields are perhaps
convenient and have a direct physical interpretation,
their use is not at all compelling, and we could have derived
exactly the same mass spectrum and the same interactions by
working in terms of meson variables.
Indeed, for $m\ne 0$, the choice between the two sets of variables
is completely irrelevant and the physical properties of the system
are exactly the same, as expected on the basis of general results
on quantum field theory. The relation between the two sets of
superfields is, however, singular at $m=0$ and the two descriptions are
no longer equivalent.

Consider, for instance, the case
$M=1$. By going back to the results of section 3, we see that
in the limit $m\to 0$ the supermultiplet $\xi^0$ becomes massless.
Indeed, if a smooth massless limit exists, we expect precisely
one massless scalar signalling the spontaneous breakdown of the global
$U(1)_R$. The other states should then be degenerate with the
Nambu--Goldstone mode because supersymmetry is unbroken.
The presence of a massless supermultiplet is confirmed by an
inspection of the scalar potential of the theory
\footnote{With the standard abuse of notation we denote with the same
symbol
the superfield and its complex lowest component.}:
\be
V=2 m^2 |F|^2 \left\vert
{\rm exp}\left({i~{\xi^0\over\sqrt{2} F}}\right)
-{\rm exp}\left(-i~{N+1\over N-1} {\xi^0\over\sqrt{2}
F}\right)\right\vert^2
\label{pot1}
\ee
For any non-vanishing value of $m$, $V$ reaches its
minimum at $\xi^0=0$. When $m$ goes to zero, $V$ vanishes identically,
leading to unbroken supersymmetry and to a complex flat direction.

Notice that the metric of the scalar manifold is given by
$1/(2\sqrt{\cm\cm^\dagger})$ in terms of mesons
and becomes singular at the minimum of the scalar potential,
when $m$ tends to zero. Such a singularity is however absent
when we look at quantities that are invariant under
reparametrizations of the fields
\footnote{I thank R. Casalbuoni for suggesting this point to me.}.
For example, the curvature of the scalar manifold vanishes at any
value of $\cm$. At the same time,
the metric associated to $\xi^0$, $|{\rm exp}(i \xi^0/\sqrt{2} F)|^2$,
is regular in the vicinity of $\xi^0=0$, for any $m\ll\la$.

We also see that in the $m\to 0$ limit, with $\la$ fixed, the decay
constant $F$ diverges, contrary to what happens in ordinary QCD.
This fact, however, does not spoil the smoothness of the limit.
Indeed, it simply means that the interaction terms, which are
suppressed by inverse powers of $F$ as shown in eqs. (\ref{ka4})
and (\ref{wexp}), go to zero. The $U(1)_R$
symmetry is broken by an ``infinite'' amount and the
low-energy theory consists of a single non-interacting chiral
superfield.
In the theory under discussion, the divergence associated to $F$ does
not affect any physical quantity. We may ask what happens if we
couple the $U(1)_R$ current to extra matter multiplets, analogously to
the case of weak interactions, where some of the left-handed
weak currents, coupled to ordinary matter, may coincide
with the chiral QCD currents. Even in this case the observable
quantities will depend on $F$ only through combinations that
remain finite in the massless limit. For instance, the
decay rate of the Nambu--Goldstone boson
into a pair of matter fields is proportional to $|F|^2 m_{\xi^0}$,
still vanishing in the $m\to 0$ limit.
The effective Lagrangian
considered here allows a small mass expansion
around the massless
limit. When $m\ll\la$, we can easily take into account the
small non-vanishing mass and the small interactions of the
chiral supermultiplet, by keeping a convenient number of terms in the
expansion of eqs. (\ref{ka4}) and (\ref{wexp}).

The general case can be discussed along similar lines.
If we perform the massless limit, keeping a fixed non-vanishing ratio
among the quark masses\footnote{If some of the masses run to zero
faster than the others, we can eliminate the ``slow'' modes
by decoupling them from the effective theory.},
we obtain exactly $M^2$ massless supermultiplets, corresponding to
the spontaneous breaking of $SU(M)\times SU(M)\times U(1)_B\times
U(1)_R$.
The transition to the massless limit is smooth, and no observable
quantity becomes singular. The formally infinite
decay constants suppress the interaction terms and the theory
experiences a free phase. When small masses are turned on, the theory
has a small departure from the free phase, with interaction terms
that can be systematically computed through an expansion in inverse
powers of the decay constants.

The Nambu--Goldstone variables occur naturally also for $M=N$.
The quantum constraint (\ref{qcon}) satisfied by the meson superfields
along the mesonic branch can be explicitly solved by:
\be
\cm=\la^2 {\rm exp}\left(i~{\xi^a T^a\over\la}\right)~~~~~,
\label{mun}
\ee
with $a=1,...,N^2-1$. More precisely, the quantum constraint
defines a complex non-linear $\sigma$-model
associated to the coset space $SL(N,C)\times SL(N,C)/SL(N,C)$.
Indeed the constraint is invariant under independent $SL(N,C)$
transformations acting from the right and from the left
on $\cm$. Moreover, the generic vacuum configuration $\langle \cm^i_j
\rangle$ can always be mapped into $\la^2 \delta^i_j$ by means
of a suitable $SL(N,C)\times SL(N,C)$ transformation, thus displaying
the invariance of the vacuum under the diagonal $SL(N,C)$
subgroup. The generic coset element leads to the above
representation for the meson variables \cite{nli}. It is interesting
to note that in the chiral limit the decay constant is
just $\la$, which coincides with a naive extrapolation of the result
(\ref{dc1}) obtained by taking $M=N$, before performing the
massless limit.

Although this physical picture is quite natural,
we should be careful before drawing any conclusion
about the massless limit for $M<N$.
When the meson variables are used, the massless limit does
not exist. The metric of the scalar manifold is singular for
infinite meson VEVs.
This singular behaviour is smoothed out
by adopting the Nambu--Goldstone superfield $\xi^a$ and
transferred to the decay constants of the theory.
Nevertheless several questions remain open.
Is there anything against using $\xi^a$ instead of $\cm$?
On what basis can we decide about the appropriate variables
for the massless case?
More generally: has the singular behaviour of mesons a
physical meaning or does it simply reflect the inadequacy
of the meson coordinates to describe the massless case?
We have not yet found a sharp test that may discriminate
between the two possibilities, and our present understanding
only suggests that a smooth massless limit of SQCD, $M<N$,
is not inconceivable.

\section{Conclusions}

In this work we have discussed in detail the properties
of SQCD, $M<N$, for masses smaller than or comparable to
the SQCD scale $\la$. To start with, we have moved from
meson variables to Nambu--Goldstone-like variables $\xi^a$.
For small quark masses, the meson superfields experience very large
fluctuations and a more convenient description of the system
is obtained by using Nambu--Goldstone superfields, which fluctuate
close to the origin in field space.
We have presented explicit expressions for the mass spectrum
and the decay constants of the system, checking that our
results satisfy the properties required by decoupling.
When the massless limit is approached, each observable
varies smoothly, and the theory approaches a free
phase, consisting of $M^2$ non-interacting chiral supermultiplets.
If the massless limit exists, as supported by rephrasing
the theory in terms of Nambu--Goldstone variables, the
free phase has a clear physical meaning: it corresponds
to the spontaneous breaking of the global flavour symmetry.
The flavour symmetry is formally broken by an infinite amount,
but this has the only effect of turning off the interactions
between the chiral superfields.
The massive case, for masses smaller than $\la$, can be
thought of as a perturbation around the massless case, with correction
terms that can be systematically computed in the low-energy theory.

The discussion outlined in this paper makes use of an
effective-Lagrangian approach. It is worth remembering that,
besides the virtues and the relative handiness of the method,
this approach has to face important limitations. These are
related to the well-known fact that
the correct identification of the elementary excitations of the
system is part of the starting assumptions, not the outcome of a
dynamical computation. Precisely for this reason,
within this approach, we have not been able to resolve
the ambiguity associated to the massless limit,
and the picture presented here should be taken as a conjecture.
It would be extremely interesting if some indications in favour
or against our proposal could come from the intensively studied
and rapidly evolving field of D-brane and M-theory constructions
where,
already now, many non-perturbative results of super-Yang--Mills
systems have been reproduced \cite{mth}.

\section*{Acknowledgements}
I have enjoyed useful discussions with L.~Alvarez--Gaum\'e,
R.~Casalbuoni, J.~P.~Hurni, C.~Kounnas, E.~Rabinovici, L.~Randall,
R.~Rattazzi and F.~Zwirner. A special thank goes to G. Veneziano
for reading the manuscript and for several comments and suggestions.
\newpage


\begin{thebibliography}{99}
\bibitem{pes}
For recent reviews see:
K. Intriligator and N. Seiberg,
Nucl. Phys. Proc. Suppl. B 45 (1996) 1 and
Nucl. Phys. Proc. Suppl. B 55 (1996) 200;\\
M. Peskin, in the Proceedings of the 1996 TASI, Boulder, Colorado, USA,
1996 (hep-th/9702094);\\
M. Shifman, lectures at the Summer School in High-Energy Physics and
Cosmology, ICTP, Trieste, Italy, 1996 (hep-ph/9704114);\\
L. Alvarez-Gaum\'e and F. Zamora, preprint CERN-TH-97-257
(hep-th/9709180).
\bibitem{comp}
T. Banks and E. Rabinovici, Nucl. Phys. B 160 (1979) 349;\\
E. Fradkin and S. Shenker, Phys. Rev. D 19 (1979) 3682;\\
S. Dimopoulos, S. Raby and L. Susskind, Nucl. Phys. B 173 (1980) 208.
\bibitem{vene}
G. Veneziano, Phys. Lett. B 128 (1983) 199;\\
G. M. Shore and G. Veneziano, Int. J. of Mod. Phys. A 1 (1986) 499.
\bibitem{tvy}
T.R. Taylor, G. Veneziano and S. Yankielowicz, Nucl. Phys. B 218 (1983)
493;\\
G. Veneziano and S. Yankielowicz, Phys. Lett. B 113 (1982) 321.
\bibitem{wit}
E. Witten, Nucl. Phys. B 202 (1982) 253.
\bibitem{buc}
F. Buccella, J.P. Derendinger, S. Ferrara and C. Savoy, Phys. Lett. 115
B (1982) 375;\\
M. Luty and W. Taylor, Phys. Rev. D 53 (1996) 3399.
\bibitem{ads}
I. Affleck, M. Dine and N. Seiberg, Nucl. Phys. B 241 (1984) 493.
\bibitem{ama}
D. Amati, K. Konishi, Y. Meurice, G.C. Rossi and G. Veneziano, Phys.
Rep. 162 (1988) 169, and references therein.
\bibitem{se1}
N. Seiberg, Phys. Rev. D 49 (1994) 6857.
\bibitem{cds}
A.C. Davis, M. Dine and N. Seiberg, Phys. Lett. B 125 (1983) 487.
\bibitem{aha}
O. Aharony, M. Peskin, J. Sonnenschein and S. Yankielowicz, Phys. Rev.
D
52 (1995)
6157;\\
S.P. Martin and J. D. Wells, hep-th/9801157.
\bibitem{gri}
M. T. Grisaru, M. Rocek and R. von Unge, Phys. Lett. B 383 (1996) 415.
\bibitem{se2}
I. Affleck, M. Dine and N. Seiberg, Nucl. Phys. B 256 (1985) 557.
\bibitem{ran}
E. Poppitz and L. Randall, Phys. Lett. B 336 (1994) 402.
\bibitem{nli}
W. Lerche, Nucl. Phys. B 238 (1984) 582;\\
W. Buchm\"uller and W. Lerche, Ann. Phys. 175 (1987) 159;\\
G. M. Shore, Nucl. Phys. B320 (1989) 202;\\
A.C.W. Kotcheff and G. M. Shore, Int. J. Mod. Phys. A 4 (1989) 4391.
\bibitem{mth}
E. Witten, Nucl. Phys. B 507 (1997) 658;\\
A. Brandhuber, N. Itzhaki, V. Kaplunovsky, J. Sonnenshein and
S. Yankielowicz, Phys. Lett. B 410 (1997) 27;\\
S. Elitzur, A. Giveon, D. Kutasov, E. Rabinovici and A. Schwimmer,
Nucl. Phys. B 505 (1997) 202;\\
K. Hori, H. Ooguri, Y. Oz, preprint LBL-40336 (hep-th/9706082).
\end{thebibliography}
\end{document}